\documentclass[reprint, amsmath, amssymb, aps,
superscriptaddress, floatfix, prd,longbibliography]{revtex4-2}

\usepackage{color}
\usepackage[dvipsnames]{xcolor}
\usepackage[normalem]{ulem}
\usepackage{float}
\usepackage{graphicx}
\usepackage{dcolumn}
\usepackage{bm}
\usepackage{braket}
\usepackage{amsmath}
\usepackage{multirow}
\DeclareMathOperator*{\argmin}{argmin}

\newcommand{\new}[1]{{\color{Black}#1}}

\begin{document}

\preprint{APS/123-QED}

\author{Joseph G. Smith}
\email{jgs46@cam.ac.uk}
\affiliation{Cavendish Laboratory, Department of Physics, University of Cambridge, Cambridge, CB3 0HE, United Kingdom}
\affiliation{Hitachi Cambridge Laboratory, J. J. Thomson Ave., Cambridge, CB3 0HE, United Kingdom}
\author{Crispin H. W. Barnes}
\email{chwb101@cam.ac.uk}
\affiliation{Cavendish Laboratory, Department of Physics, University of Cambridge, Cambridge, CB3 0HE, United Kingdom}
\author{David R. M. Arvidsson-Shukur}
\email{drma2@cam.ac.uk}
\affiliation{Hitachi Cambridge Laboratory, J. J. Thomson Ave., Cambridge, CB3 0HE, United Kingdom}

\title{Risk-minimizing states for the quantum-phase-estimation algorithm}

\begin{abstract}

The quantum-phase-estimation algorithm (QPEA) is widely used to find estimates of  unknown phases.
The original algorithm relied on an input state in a uniform superposition of all possible bit strings. However, it is known that other input states can reduce certain Bayesian risks of the final estimate.
Here, we derive a method to find the risk-minimizing input state for any risk.
These states are represented by an eigenvector of a Toeplitz matrix with elements given by the Fourier coefficients of the loss function of interest.
We show that, while the true optimal state does not have a closed form for a general loss function, it is well approximated by a state with a cosine form.
When the cosine frequency is chosen appropriately, these states outperform the original QPEA and achieve the optimal theoretical quantum-advantage scaling for three common risks.
Furthermore, we prove that the uniform input state is  suboptimal for any reasonable loss function.
Finally, we design methods to mitigate the impact of depolarizing noise on the performance of QPEA.

\end{abstract}

\maketitle

\section{Introduction}

Quantum phase estimation is an important subroutine for quantum technologies. It is used in several contenders for nonclassical advantage. 
Examples of such contenders are quantum algorithms \cite{QAE, QAE2, dave2, shor, dave1}, quantum chemistry \cite{algos1, VQE, VQE2}, and interferometry \cite{inter2, clocks, grav_waves}. 
A vital component in these examples is the estimation of an unknown phase $\theta$ applied via a unitary evolution $U(\theta)$. Due to the random nature of quantum measurements, estimates of $\theta$, $\hat{\theta}$, will have an intrinsic variance, and hence an error. 
The goal of a phase-estimation algorithm is to minimize the error of $\hat{\theta}$ with respect to some available resource, typically how many times $U(\theta)$ is applied across the algorithm. 
The use of entanglement or coherence can achieve a theoretical quadratic improvement of estimator variance \cite{Lloyd1, Lloyd2}.
In particular, for $N$ applications of the unknown unitary $U(\theta)$, a classically shot-noise-limited \new{estimate's variance} scales as $1/N$, while a nonclassically Heisenberg-limited estimate's \new{variance} scales as $1/N^2$.
However, few algorithms achieve this quadratic advantage due to issues with a lack of point-identification (the output quantum state or measurement data could arise from multiple distinct values of the unknown parameter $\theta$) \cite{me} or decoherence in the quantum system \cite{noisy_QFI, noisy_QFI2}.

A point-identified estimate of $\theta$ can be found by measuring several different circuits and using classical post-processing on the observed outcomes, either relying on Bayesian inference or a classical discrete Fourier transform \cite{me2, berry, fourier, fourier2, multiphase}. Thus, one can rule out all-but-one of the potential estimates of $\theta$.
Alternatively, one can achieve point identification via \new{a measurement of} a single quantum circuit \new{that uses} the inverse quantum Fourier transform, \new{implemented} either using two-qubit gates \cite{nielsen} or feedback-based methods \cite{semiQFT, semiQFT2}. 
The resulting, fully quantum algorithm is often known as the quantum-phase-estimation algorithm (QPEA) \cite{QPEA, QPEA2, QPEA_vs_MLE}.
The QPEA is one of the most used subroutines in the field of quantum algorithms.
The traditional QPEA circuit is initialized with a register of $m$ qubits in an equal mix of the computational basis states. 
The error of the obtained estimate is often quoted as a Heisenberg-limited error because the QPEA gives a correct binary representation of ${\theta}/{2\pi}$ with probability greater than $40\%$ \cite{nielsen}.
However, this error is deceiving, as one cannot know if the algorithm has succeeded without first knowing the true value of $\theta$.
From a complexity-theoretic perspective, meaningful errors should consider the errors from all outcomes. An example of such an error measure is the Bayes risk.
The Bayes risk is the average of the expected posterior loss over all possible measurement outcomes with respective to some loss function.
Previous works have shown that the QPEA does not achieve the quadratic (Heisenberg-limited) improvement for risks calculated using the Holevo variance \cite{higgins, berry}. 
However, a quantum advantage can be reinstated by initializing the register of the same circuit in a different state than the initially-suggested uniform-superposition state \cite{berry, opt_states, cosine_windows}.

In this Article, we expand on previous works by constructing a methodology to find the state that the QPEA register should be prepared in to minimize a general risk.
We find that the optimal state can be represented by the eigenvector of a Toeplitz matrix with elements determined by the Fourier coefficients of the loss function used to calculate the risk.
We use this fact to prove that the traditional QPEA state only minimizes risks calculated using a trivial constant loss functions, a situation where every measurement process, even a random guess of $\theta$, has the same risk.
For general loss functions, we show that, while the optimal state does not have a closed form, such states are well approximated by a state whose amplitudes have a cosine form over the computational basis.
By tuning the frequency of the cosine envelope, one can decrease the risk to values below that of the traditional QPEA. 
For risks calculated using the Holevo variance, our approximate states give results akin to previously derived optima \cite{opt_states}.
We further demonstrate that the cosine states can obtain a quadratic improvement of three other common loss functions when compared to the shot-noise limits of those risks.
The traditional QPEA cannot.
Finally, we show that, even in the presence of depolarizing noise, risks better than the shot-noise limit can be obtained by repetitive measurements of circuits initialized with cosine states.

\section{Optimal QPEA register states}

\begin{figure}
    \centering
    \includegraphics[width=\linewidth]{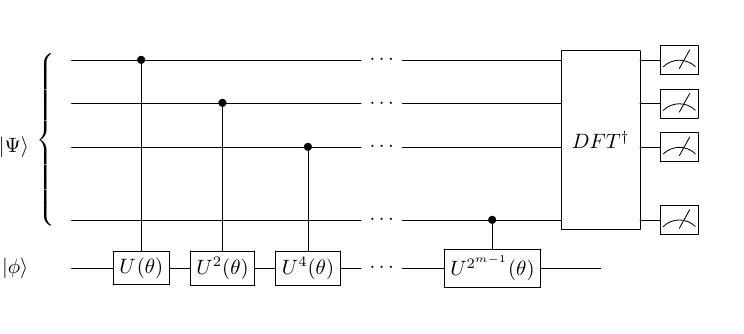}
    \caption{The QPEA circuit.}
    \label{fig:circuit}
\end{figure}

In quantum phase estimation one estimates an unknown phase $\theta$ of a unitary $U(\theta)$. To do so, one uses an eigenstate $\ket{\phi}$ such that $U(\theta)\ket{\phi} = e^{i \theta} \ket{\phi}$. 
For this state, the phases $\theta + 2 \pi l$ for integer $l$ are equivalent, so unique unitaries will have $\theta \in \Theta=[\theta_0-\pi,\theta_0+\pi]$.
\new{Here, $\theta_0$ is the center of the interval $\Theta$, commonly set to either $0$ or $\pi$.}
The QPEA circuit is shown in Fig. \ref{fig:circuit}.
The input state is spread over two registers.
The first register consists of $m$ qubits initially prepared in the state 
\begin{equation}
    \label{eqn:PSI}
    \ket{\Psi} = \sum_{j=0}^N c_j \ket{j},
\end{equation}
where $N=2^m-1$ and $c_j \in \mathbb{R}$.
The second register is prepared in the state $\ket{\phi}$.
The $i^{\textrm{th}}$ qubit of the first register is used as the control of $2^{i-1}$ $\textrm{controlled-}U(\theta)$ gates with the second register as the target.
In total, $N$ $\textrm{controlled-}U(\theta)$ gates are applied in the circuit. Throughout this work, we shall consider \new{the total number of $U(\theta)$ gates used} as the computational resource. A final measurement of the first register, yields a bit-wise estimate of $\theta/2 \pi$.

In this work, we shall also consider the effect of applying some noise to the quantum state throughout the circuit.
We adopt a simple and analytically-tractable noise model in which the first register is subject to a depolarizing channel of strength $\lambda$ after each $U(\theta)$ gate.
The final state of the first register at the time of measurement with this noise is
\begin{equation}
    \rho = (1-\lambda)^{N} \ket{\Phi} \! \! \bra{\Phi} + \frac{1-(1-\lambda)^{N}}{2^m} \cdot I ,
\end{equation}
where
\begin{equation}
    \ket{\Phi} = \frac{1}{\sqrt{2^m}} \sum_{j=0}^N \sum_{k=0}^N c_k e^{ik\left( \theta - \frac{2 \pi j}{2^m}\right)}\ket{j}.
\end{equation}
The probability that the first register is measured in the state $\ket{y}$, i.e., the likelihood function, is
\begin{equation}
\label{eqn:p_y}
\begin{split}
    p(y|\theta) &= \frac{1}{2^m} + \frac{(1-\lambda)^N}{2^m} \sum_{\substack{k=-N \\ k \neq 0}}^N \sum_{j=0}^N c_j c_{j-k} e^{ik\left(\theta - \frac{2 \pi y}{2^m} \right)}
\end{split} .
\end{equation}
After the first register is measured in the state $\ket{y}$, a posterior distribution $p(\theta|y)$ can be generated using Bayes' rule:
\begin{equation}
\label{eqn:bayes}
    p(\theta|y) = {\pi(\theta)p(y|\theta)}/{p(y)}.
\end{equation}
Here, $\pi(\theta)$ is the prior distribution and 
\begin{equation}
    p(y) = \int_\Theta \pi(\theta)p(y|\theta) d\theta.
\end{equation}
The prior distribution of most interest is the uniform prior, $\pi(\theta)=1/2\pi$, which represents a complete lack of initial knowledge of the value of $\theta$ before any measurement is performed.
For a uniform prior, $p(y)=1/2^m$ for any initial state $\ket{\Psi}$.
$p(\theta|y)$ allows one to construct an expected value of a loss function for any estimate $\hat{\theta}'$:
\begin{equation}
\label{eqn:PEL}
    \mathcal{L}(y,\hat{\theta}') = \int_\Theta p(\theta|y) L(\hat{\theta}',\theta) d\theta,
\end{equation}
where $L(\hat{\theta}',\theta)$ is a loss function.
The estimate that we make is given by minimizing the expected loss:
\begin{equation}
    \hat{\theta} = \argmin_{\hat{\theta}'} \mathcal{L}(y,\hat{\theta}') .
\end{equation}
The expected value of $\mathcal{L}(y,\hat{\theta})$ \new{for a given register state $\ket{\Psi}$} over all outcomes $y$ is known as the risk:
\begin{equation}
\label{eqn:risk}
    \mathcal{R}(\Psi) = \sum_y p(y) \mathcal{L}(y,\hat{\theta}) = \sum_y \int_\Theta \pi(\theta)p(y|\theta) L(\hat{\theta},\theta) d\theta.
\end{equation}
Our work aims to establish a methodology that, for a given loss function and amount of resources, allows one to \new{find the optimal state  $\ket{\Psi'}$ that minimizes $\mathcal{R}(\Psi)$ over all initial-register states $\ket{\Psi}$.}

Loss functions of particular interest are of the form $L(\hat{\theta},\theta)=L(\delta)$, where $\delta = \theta - \hat{\theta}$ is the distance between $\theta$ and $\hat{\theta}$.
In this paper, we restrict our analysis to loss functions that are non-decreasing with increasing $|\delta|$ and satisfy $L(\delta)=L(-\delta)$.
In this case, $\mathcal{L}(y,\hat{\theta})$ measures the average distance between $\hat{\theta}$ and possible true values of $\theta$.
The four loss functions we study are summarized in Table \ref{tab:l_k}.
While three of the loss functions are commonly studied, the $1$-$0$ loss function is not. 
\new{The $1$-$0$ loss is one if $|\delta|>\epsilon$ and zero otherwise, where $\epsilon>0$.
In this case, $\mathcal{R}(\Psi)=\textrm{Pr} [|\hat{\theta}-\theta|>\epsilon ]$.}
%$\mathcal{R}(\Psi)$ is the average probability that there is at least an $\epsilon$ difference between the final estimate $\hat{\theta}$ and the true value of $\theta$.
Due to equivalences of the phases $\theta + 2\pi l$ for integer $l$, every loss function should be periodic with period $2\pi$.
A non-periodic loss function $L_{\textrm{NP}}(\delta)$ of the above form can be converted into a periodic function $L(\delta)$ by calculating
\begin{equation}
    L(\delta) = \min_{l \in \mathbb{Z}} \left\{ L_{\textrm{NP}}(\delta+2\pi l)\right\}.
\end{equation}
If $\theta_0=\hat{\theta}$ in $\Theta$, $\delta \in[-\pi,\pi]$. 
One finds that $L(\delta) = L_{\textrm{NP}}(\delta)$, so this conversion can be ignored.
Periodicity allows the loss function to be expanded as the Fourier series $L(\delta) = \sum_k L_k e^{ik\delta}$, where
\begin{equation}
\begin{split}
\label{eqn:coeffs}
    L_k &= \frac{1}{2\pi} \int_{-\pi}^{\pi} L(\delta) e^{-ik\delta} d\delta.
\end{split}
\end{equation}
The Fourier coefficients satisfy both $L_{-k} = L_{k}^*$ for real loss functions and
$L_k \in \mathbb{R}$ for even loss functions.

\begin{table}[]
    \centering
    \begin{tabular}{|c|c|c|}
        \hline
        \multirow{2}{*}{Loss} & \multirow{2}{*}{$L(\delta)$} & \multirow{2}{*}{Non-zero Fourier coefficients} \\
        & & \\
        \hline 
        \hline 
        Absolute & $L(\delta)=|\delta|$ & $L_k = \begin{cases}
            \pi / 2 & k = 0\\
            -2/\pi k^2 & k = \textrm{odd}
        \end{cases}$ \\ 
        \hline
        Squared & $L(\delta)=\delta^2$ & $L_k = \begin{cases}
            \pi^2 / 3 & k = 0\\
            2(-1)^k/k^2 & k \neq 0
        \end{cases}$ \\
        \hline
        Holevo & $L(\delta)=4 \sin^2(\delta/2)$ & $L_k = \begin{cases}
            2 & k = 0\\
            -1 & |k|=1
        \end{cases}$ \\
        \hline
        $1$-$0$ & $L(\delta)= \begin{cases}
            1 & |\delta| > \epsilon \\
            0 & |\delta| \leq \epsilon
        \end{cases}$ & $L_k = \begin{cases}
            1-\epsilon/\pi & k = 0\\
            -\sin(k\epsilon)/\pi k & k \neq 0
        \end{cases}$ \\
        \hline
    \end{tabular}
    \caption{The Fourier coefficients of different loss functions, where $\delta = \theta - \hat{\theta}$.}
    \label{tab:l_k}
\end{table}

In terms of these Fourier coefficients, Eq. \eqref{eqn:PEL} becomes
\begin{equation}
\begin{split}
    \mathcal{L}(y,\hat{\theta}') = L_0 + (1-\lambda)^N \sum_{k=1}^N \sum_{j=k}^N c_j c_{j-k} L_k \cos k \vartheta,
\end{split}
\end{equation}
where $\vartheta=\hat{\theta}' - {2 \pi y}/{2^m}$.
This loss is minimised by the estimate $\hat{\theta}=2\pi y/2^m$:
\begin{equation}
    \mathcal{L}(y,\hat{\theta}) = L_0 + (1-\lambda)^N \sum_{k=1}^N \sum_{j=k}^N c_j c_{j-k} L_k.
\end{equation}
$\mathcal{L}(y,\hat{\theta})$ is independent of the observed value of $y$.
Equation \eqref{eqn:risk} shows that the risk of \new{the QPEA with a register prepared in the state $\ket{\Psi}$ is}
\begin{equation}
\label{eqn:risk_sum}
        \mathcal{R}(\Psi)=L_0 + (1-\lambda)^N \sum_{k=1}^N \sum_{j=k}^N c_j c_{j-k} L_k.
\end{equation}

If depolarizing noise is present in each $U(\theta)$ gate ($\lambda>0$), $\mathcal{R}(\Psi)\to L_0$ for large $N$ for any initial register state.
Therefore, the risk of estimates made using the QPEA \new{with a register prepared in any state} tends to a constant value: 
$\mathcal{R}(\Psi)=O(1)$.

The risk-minimizing state can be found by re-writing Eq. \eqref{eqn:risk_sum} into the matrix equation 
\begin{equation}
    \mathcal{R}(\Psi)=L_0 + (1-\lambda)^N \boldsymbol{c}^{\dagger} \boldsymbol{R} \boldsymbol{c}.
\end{equation}
Here, $\boldsymbol{c}$ is a column vector of the values $c_j$ and
\begin{equation}
    \label{eqn:risk_matrix}
    \boldsymbol{R}=\begin{pmatrix}
        0 & L_{-1} & L_{-2} & \cdots & L_{-N} \\
        L_1 & 0 & L_{-1} & \cdots & L_{1-N} \\
        L_2 & L_1 & 0 & \cdots & L_{2-N} \\
        \vdots & \vdots & \vdots & \ddots & \vdots \\
        L_{N} & L_{N-1} & L_{N-2} & \cdots & 0
    \end{pmatrix}.
\end{equation}
For even loss functions, $\boldsymbol{R}$ is a symmetric banded Toeplitz matrix.
\new{The optimal register state $\ket{\Psi'}$ with a minimal risk of $\mathcal{R}(\Psi')$ is represented by the column vector $\boldsymbol{c}$ that is an eigenvector of $\boldsymbol{R}$ with the smallest eigenvalue. 
$\ket{\Psi'}$ is independent of $\lambda$. }

When risk is determined through the Holevo variance $L(\delta)=4 \sin^2 (\delta / 2)$, $\boldsymbol{R}$ is a tridiagonal matrix with the minimum eigenvector described by
\begin{equation}
\label{eqn:h_cos_state}
    c_i(\omega) = \sqrt{\frac{2}{N+2}} \cdot \cos\left[ \left(\frac{N}{2}-i \right) \cdot \frac{\pi}{N+2} \right].
\end{equation}
This result is in agreement with the state proposed in Ref. \cite{opt_states}. 
The minimal Holevo risk with this state is 
\begin{equation}
    \mathcal{R}(\Psi') = 2-2 (1-\lambda)^N \cos \left( \frac{\pi}{N+2}\right).
\end{equation}
Unfortunately, no closed-form expression exists for the minimum-eigenvalue eigenvector of general banded Toeplitz matrices \cite{toeplitz2, toeplitz}.
Therefore, numerical methods are needed to find the optimal state for other risks, such as the absolute, squared and $1$-$0$ risks.

One can use the above analysis to solve an alternative task:
Find the initial state $\ket{\Psi}$ required to prepare the $m$-qubit output register in the state $\ket{\left\lfloor 2^m \theta /2\pi \right\rceil}$ with the highest probability. This output state is an $m$-bit binary representation of $\theta/2\pi$.
The preparation of $\ket{\left\lfloor 2^m \theta /2\pi \right\rceil}$ is integral to many quantum algorithms, such as Shor's algorithm \cite{shor} and the HHL algorithm \cite{hhl}.
The task is equivalent to finding the state that minimizes the probability that $|\hat{\theta}-\theta|\geq \pi/2^m$ averaged over all $\theta$, or, using the terminology above, the risk calculated using the $1$-$0$ loss function with $\epsilon=\pi/2^m$.
Therefore, the optimal state for the desired preparation is the eigenstate with the minimal eigenvalue of the matrix in Eq. \eqref{eqn:risk_matrix} with $L_k=\sin(k\pi/2^m)/\pi k$. 
%Next, we investigate the form of this optimal state. 

\section{Cosine states}

\begin{figure}
    \centering
    \includegraphics[width=0.75\linewidth]{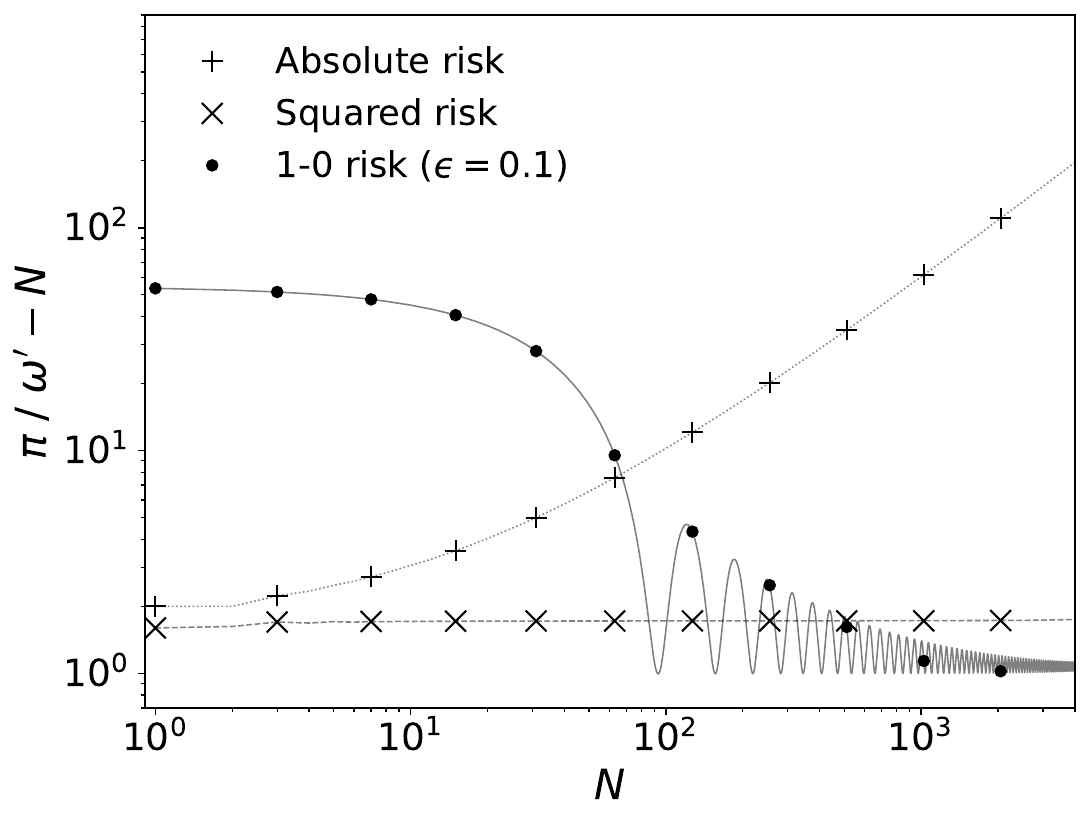}
    \caption{Numerically calculated $\pi/\omega'-N$ for different values of $N$. 
    \new{The y-axis function is chosen to aid visualization.}
    The markers plot data for integer values of $m$ with the lines plotting calculated data for non-integer values of $m$.
    For large $N$, linear regression shows that $\omega'$ scales as $\frac{\pi}{1.04N + 22.27}$, $\frac{\pi}{N + 1.72}$ and $\frac{\pi}{N + 1.00}$ for the absolute, squared and $1$-$0$ loss functions, respectively.}
    \label{fig:omegas}
\end{figure}

Instead of calculating the optimal state numerically and finding a state that may be difficult to prepare in practice, we find an approximation to these optimal states.
The symmetry of the rows of $\boldsymbol{R}$ impose two conditions on the minimum eigenvector: $c_i=c_{N-i}$ and $c_0 < c_1 < \dots < c_{\frac{N-1}{2}}$.
These conditions are satisfied by the approximation
\begin{equation}
    c_i(\omega) = \sqrt{\frac{2 \sin \omega}{2^m \sin \omega + \sin (2^m\omega)}} \cdot \cos\left[ \left(\frac{N}{2}-i \right) \omega \right].
\end{equation}
We refer to the initial state with these coefficients  as the cosine state with frequency $\omega$.
Notice that the state in Eq. \eqref{eqn:h_cos_state} is a cosine state with $\omega=\pi/(N+2)$.
When using a cosine state, the risk of $\hat{\theta}={2 \pi y}/{2^m}$ is
\begin{equation}
\label{eqn:cos_risk}
\mathcal{R}(\omega) = L_0+2(1-\lambda)^N\sum_{k=1}^N L_k f_k,
\end{equation}
where
\begin{equation}
    f_k = \frac{(2^m-k) \cos k \omega \sin \omega + \sin \left[\left(2^m-k\right) \omega \right]}{2^m {\sin \omega} + {\sin \left(2^m \omega \right)}} .
\end{equation}
The optimal cosine state has a frequency of $\omega' = \argmin_\omega \mathcal{R}(\omega)$ and a risk of $\mathcal{R}(\omega')$.
We show numerically calculated values of $\omega'$ for the absolute, squared and $1$-$0$ loss functions in Fig. \ref{fig:omegas}.

\begin{figure}
    \centering
    \includegraphics[width=\linewidth]{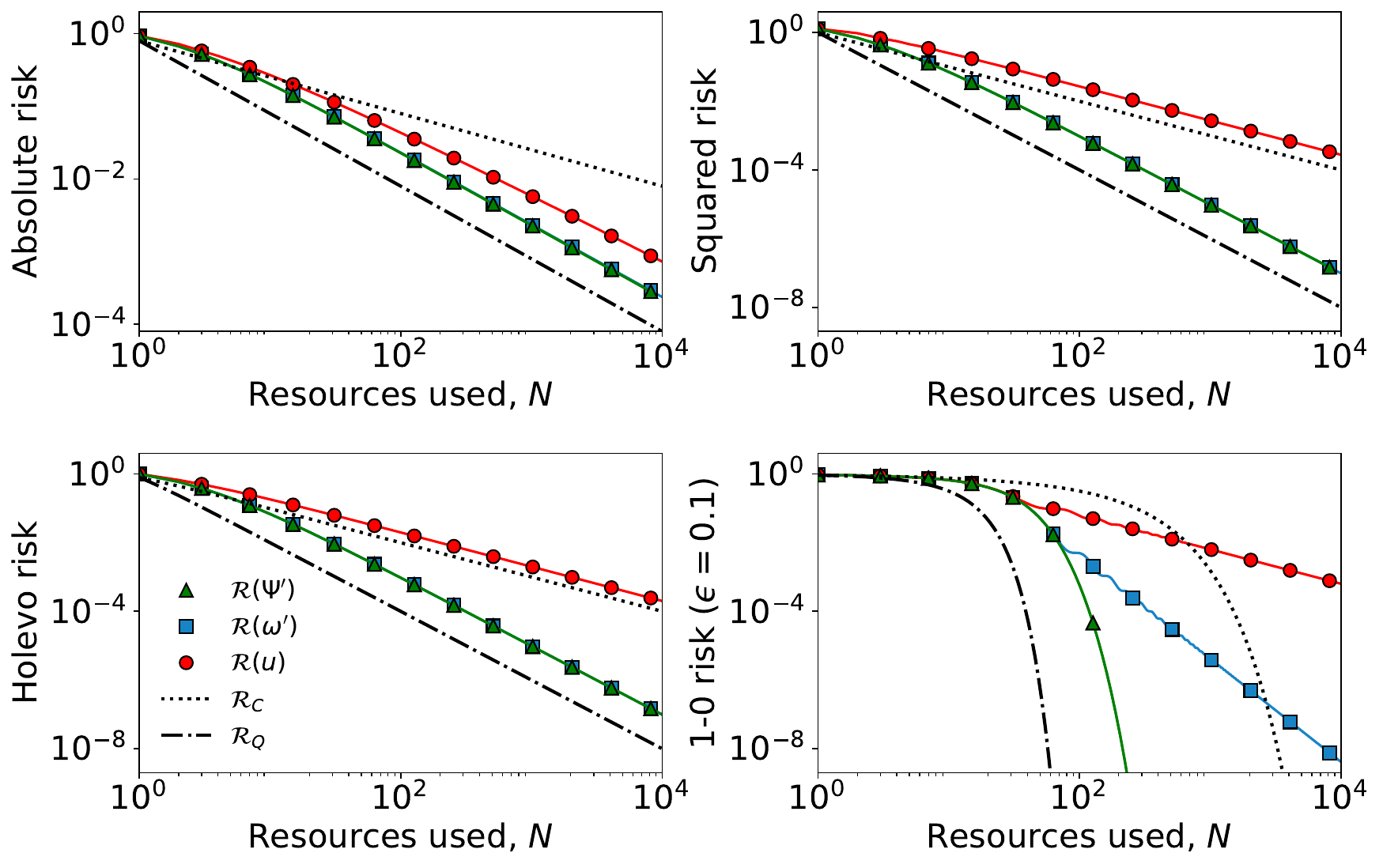}
    \caption{The four common risks from measurements of a noiseless QPEA circuit with a register prepared in different states.
    $\mathcal{R}_{\textrm{C}}$ and $\mathcal{R}_{\textrm{Q}}$ are computed using $\sigma^2=1/N$ and $\sigma^2=1/N^2$, respectively.}
    \label{fig:noiseless}
\end{figure}

Table \ref{tab:risk_scalings} summarizes the noise-free ($\lambda=0$) risk scaling of the absolute, squared, Holevo and $1$-$0$ losses using the optimal cosine states in the large-$N$ limit.
These risks are plotted for finite $N$ in Fig. \ref{fig:noiseless},
alongside $\mathcal{R}(\Psi')$.
$\mathcal{R}(\omega') \approx \mathcal{R}(\Psi')$ for the absolute, squared and Holevo risks.
\new{$\mathcal{R}(\omega') \approx \mathcal{R}(\Psi')$ is also seen for the $1$-$0$ risk when $m<7$, although $\mathcal{R}(\omega') > \mathcal{R}(\Psi')$ when $m \geq 7$.}
This divergence is due to the cosine approximation of $\ket{\Psi'}$ breaking down because $\ket{\Psi'}$ takes roughly a normal-distribution-like form over the computational basis states.

\begin{figure}
    \centering
    \includegraphics[width=\linewidth]{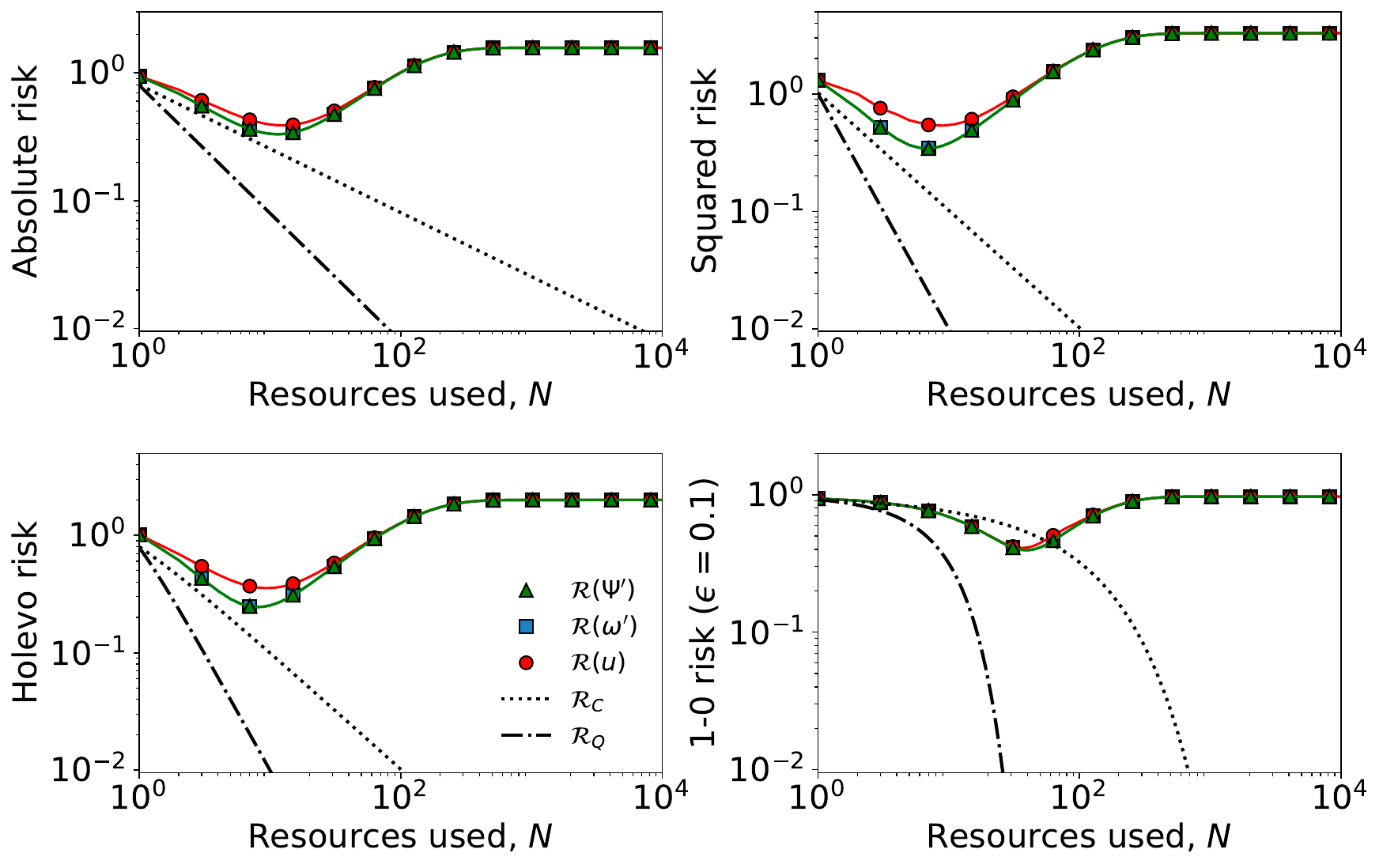}
    \caption{The four common risks from measurements of a QPEA circuit with $\lambda=1\%$.
    $\mathcal{R}_{\textrm{C}}$ and $\mathcal{R}_{\textrm{Q}}$ are computed using $\sigma^2=1/N$ and $\sigma^2=1/N^2$, respectively.}
    \label{fig:noise}
\end{figure}

Additionally, the corresponding risks in systems with $\lambda=1\%$ are plotted in Fig. \ref{fig:noise}.
\new{One observes that $\mathcal{R}(\omega') \approx \mathcal{R}(\Psi')$ for all risks.
As stated above, when depolarizing noise is present, these risks tend to $L_0$.
A minimum risk is seen for small $N$.
As $\lambda$ decreases, this achievable minimum risk also decreases. }

The above risks can be compared to the theoretically optimal scalings achieved by Fisher-information-maximizing measurements \cite{Lloyd1, Lloyd2}.
These measurements consist of a number $\nu$ measurements of a single circuit that apply $n$ copies of $U(\theta)$ to a single qubit coherently.
See Ref. \cite{me2} for more details.
Estimates of $\theta$ achieved by these measurements tend to the normal distribution $N(\theta,\sigma^2)$ for large $\nu$, where \new{$\sigma^2$ is the variance of the distribution}:
$\sigma^{-2} = {(1-\lambda)^{2n}n^2 \nu }$.
As a function of the total resource use $N=n \times \nu$, one finds that $\sigma^2=O(1/N)$ if $\nu \propto N$ and $n$ is constant.
This is the shot-noise-limited variance seen in classical statistical inference \cite{Lloyd1}.
If $n \propto N$ and $\nu$ is constant instead, $\sigma^2=O(1/N^2)$ when depolarizing noise is absent.
This is the Heisenberg limit, a quadratic quantum advantage \cite{Lloyd1}.
\new{Difficulties with point-identification prevent the Heisenberg limit from being obtained exactly in practice \cite{me}, so this limit is often used as a theoretical best-case scaling limit only.}

These variance limits can be used to calculate equivalent limits in risks by taking an expected value of a loss function over the corresponding normal distribution:
\begin{equation}
\label{eqn:classic_risk}
\begin{split}
    \mathcal{R} &= \frac{1}{\sqrt{2 \pi \sigma^2}} \int_{-\infty}^{\infty} e^{-\delta^2/2\sigma^2} L(\delta) d \delta.
\end{split}
\end{equation}
Risks calculated using a shot-noise scaling $\sigma^2$, $\mathcal{R}_{\textrm{C}}$, and a Heisenberg limited $\sigma^2$, $\mathcal{R}_{\textrm{Q}}$, are presented in Table \ref{tab:risk_scalings}.
For risks calculated using the absolute, squared and Holevo loss functions, the scalings of $\mathcal{R}(\omega')$ and $\mathcal{R}_{\mathcal{Q}}$ are equal when depolarization is absent, indicating quantum advantage.
For risks calculated using the $1$-$0$ loss function, $\mathcal{R}(\omega')$ scales worse than $\mathcal{R}_{\mathcal{C}}$.
The same is true for all the risks when depolarizing noise is present.

\begin{table}[]
    \centering
    \begin{tabular}{|c||c|c|c|c|}
        \hline
        \multirow{2}{*}{Loss} & \multirow{2}{*}{$\mathcal{R}_{\textrm{C}}$} & \multirow{2}{*}{$\mathcal{R}_{\textrm{Q}}$}& \multirow{2}{*}{$\mathcal{R}(u)$} & \multirow{2}{*}{$\mathcal{R}(\omega')$} \\
        & & & & \\
        \hline 
        \hline 
        Absolute & \multirow{2}{*}{$O\left(\frac{1}{\sqrt{N}}\right)$} & \multirow{2}{*}{$O\left(\frac{1}{{N}}\right)$} & \multirow{2}{*}{$O\left(\frac{\ln N}{{N}}\right)$} & \multirow{2}{*}{$O\left(\frac{1}{{N}}\right)$} \\ 
        loss & & & & \\
        \hline
        Squared & \multirow{2}{*}{$O\left(\frac{1}{{N}}\right)$} & \multirow{2}{*}{$O\left(\frac{1}{{N^2}}\right)$} & \multirow{2}{*}{$O\left(\frac{1}{{N}}\right)$} & \multirow{2}{*}{$O\left(\frac{1}{{N}^2}\right)$} \\
        loss & & & & \\
        \hline
        Holevo & \multirow{2}{*}{$O\left(\frac{1}{{N}}\right)$} & \multirow{2}{*}{$O\left(\frac{1}{{N^2}}\right)$} & \multirow{2}{*}{$O\left(\frac{1}{{N}}\right)$} & \multirow{2}{*}{$O\left(\frac{1}{{N}^2}\right)$} \\
        variance & & & & \\
        \hline
        $1$-$0$ & \multirow{2}{*}{$O\left(e^{-N\epsilon/2}\right)$} &  \multirow{2}{*}{$O\left(e^{-N^2\epsilon/2}\right)$} & \multirow{2}{*}{$O\left(\frac{1}{{N}\epsilon}\right)$} & \multirow{2}{*}{$O\left(\frac{1}{{N}^3\epsilon}\right)$} \\
        loss & & & & \\
        \hline
    \end{tabular}
    \caption{The large-$N$ scalings of different risks in noiseless systems.}
    \label{tab:risk_scalings}
\end{table}

A second comparison can be made to measurements with the uniform-superposition state used by the traditional QPEA: $c_i=2^{-m/2}$ \cite{QPEA, nielsen}.
The uniform state is equal to a cosine state with $\omega=0$.
The risk of estimates with this state is
\begin{equation}
\label{eqn:uniform_risk}
    \mathcal{R}(u) = L_0 + 2 (1-\lambda)^N \sum_{k=1}^{N} L_k \left(1 - \frac{k}{2^m} \right).
\end{equation}
Because the minimization to find $\omega'$ is taken over all $\omega$ including $\omega=0$, $\mathcal{R}(\omega')\leq \mathcal{R}(u)$.
Therefore, estimates made with the optimal cosine state has an equal or smaller risk than estimates made with the uniform state.

Evaluating Eq. \eqref{eqn:uniform_risk} for noiseless circuits with either the absolute, squared, Holevo or $1$-$0$ loss functions gives
\begin{equation}
\label{eqn:four_uniform_qpea_risks}
    \begin{split}
        \mathcal{R}(u) &= \frac{\ln 4 + \psi^{(0)} \left( \frac{N+2}{2}\right) + \gamma_C }{\pi 2^{m-1}} + \frac{\psi^{(1)} \left( \frac{N+2}{2}\right)}{\pi}, \\
        \mathcal{R}(u) &= \frac{\ln 2 + \Phi(2^m)}{2^{m-2}} + \psi^{(1)} \left( \frac{2^m+1}{2}\right) - \psi^{(1)} \left( 2^{m-1} \right),  \\
        \mathcal{R}(u) &= \frac{2}{2^m}, \\
        \mathcal{R}(u) &= \frac{\sin \left( \frac{N \epsilon}{2} \right)\sin \left( {2^{m-1}\epsilon} \right)}{\pi 2^{m-1} \sin \left( \frac{\epsilon}{2} \right)} + 2\sum_{k=0}^{\infty} \frac{ \sin\left[(k+2^m)\epsilon \right]}{\pi(k+2^m)},
    \end{split}
\end{equation}
respectively. Here, $\psi^{(n)}(z)$ is the $n^{\textrm{th}}$ derivative of the digamma function, $\gamma_C$ is Euler's constant and $\Phi(a) = \int_{t=0}^{\infty} \frac{e^{-at}}{1+e^{-t}} dt$.
For large $N$, these risks tend to the limits displayed in Table \ref{tab:risk_scalings}.
These four risks scale worse than $\mathcal{R}_{\mathcal{Q}}$. 
Only the absolute risk scales better than $\mathcal{R}_{\mathcal{C}}$.
All of these risks tend to $L_0$ if $\lambda>0$ for the reasons given above.

One may ask when the uniform state is optimal.
This optimality occurs when the column vector of ones is the minimum eigenvector of $\boldsymbol{R}$.
\new{When the register consists of one qubit ($m=1$)}, this column vector is the minimum eigenvector when $L_1 \leq 0$, which is satisfied by any even non-decreasing loss function.
For optimality at a given register size $m$, one requires that $L_{j} = L_{2^m-j}$ for positive integer $j$.
For this condition to hold for every $m$, all coefficients $L_j$ for $j>0$ must be equal.
However, a finite loss function has $L_N \to 0$ for large $N$ due to the decay of Fourier coefficients. 
Thus, $L_j = 0$ for all $j>0$.
Loss functions that satisfy this condition are of the form $L(\delta)=L_0$, a trivial constant. 
With this loss function, Eq. \eqref{eqn:risk} suggests that $\mathcal{R}(\Psi) = L_0$ for any initial register state.
\new{In fact, the risk is $L_0$ for any estimation procedure, even a random guess of $\theta$}.
Therefore, one concludes that estimation with the uniform state is optimal for any $m$ only in pathological situations where the QPEA does not do better in determining $\theta$ than random guesses.

\section{Multiple measurements of a single circuit}

\begin{figure*}
    \centering
    \includegraphics[width=0.8\linewidth]{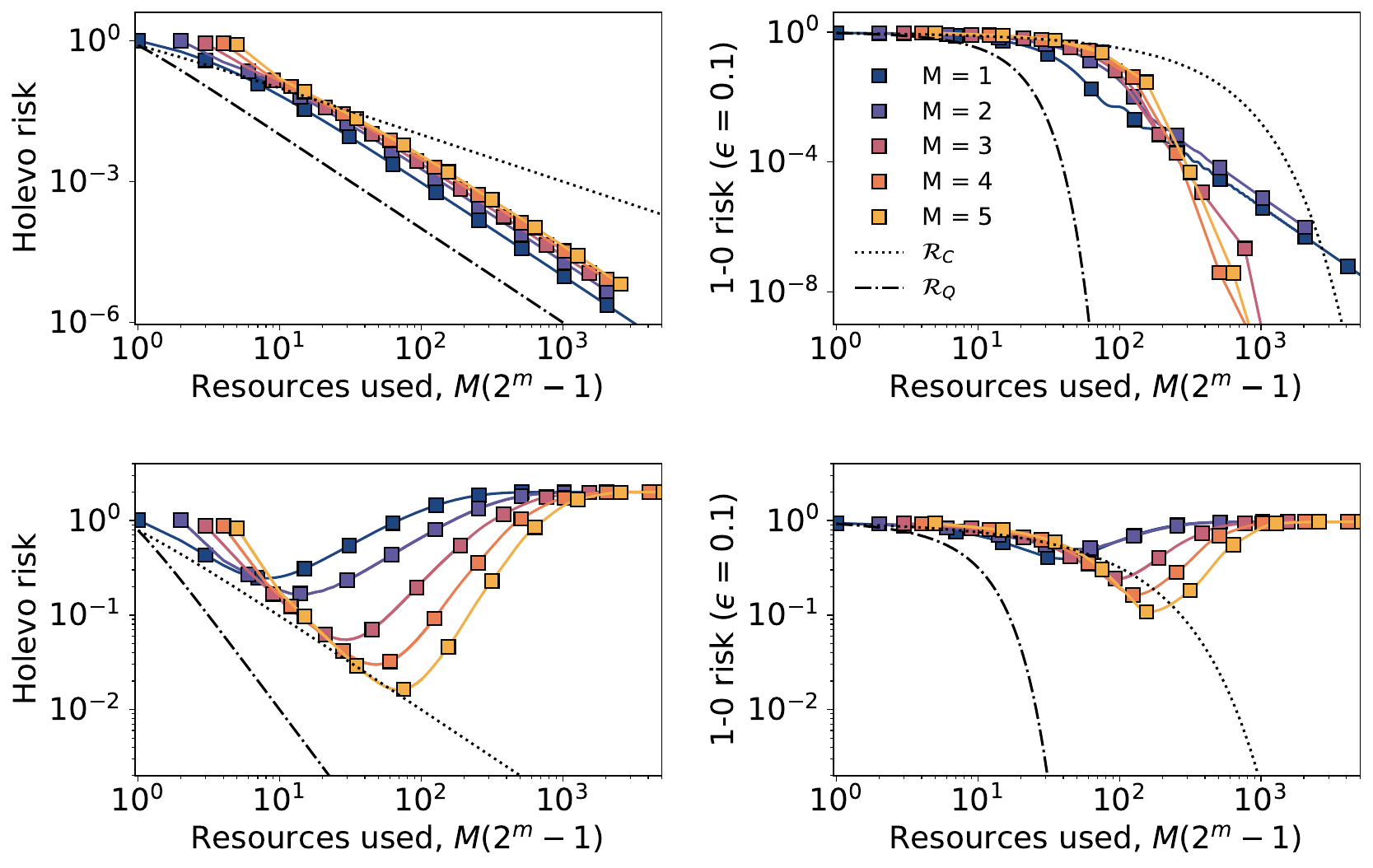}
    \caption{\new{The risk of estimates of $\theta$ made by $M$ measurements of the QPEA with an $m$-qubit register prepared in the optimal cosine state against the total resource use $M(2^m-1)$.
    The top row shows risks when using noiseless circuits, whilst the bottom row shows risks when using circuits with $\lambda = 1\%$.}}
    \label{fig:mm}
\end{figure*}

The performance of the QPEA is limited by the fact that resources can only be added to the algorithm by increasing the size of the input register, and thus circuit depth. 
We now consider a simple modification to the QPEA. In particular, we consider making multiple evaluations of the same QPEA circuit (with a fixed register size $m$) and then analyzing the results. This provides a method to increase resources without increasing the level of depolarization present in the algorithm.
Consider carrying out a number $M$ measurements of the QPEA circuit in Fig. \ref{fig:circuit}.
\new{In total, $M(2^m-1)$ applications of $U(\theta)$ are used during this protocol.}
The likelihood of observing the outcome $\boldsymbol{y}=(y_1,y_2,\dots,y_M)$ is
\begin{equation}
\begin{split}
    p(\boldsymbol{y}|\theta) &= \prod_{i=1}^M p(y_i|\theta).
\end{split}
\end{equation}
This new likelihood function $p(\boldsymbol{y}|\theta)$ can be substituted for $p(y|\theta)$ in Eqs. \eqref{eqn:bayes}-\eqref{eqn:risk} to find the measurement risk.
No closed-form solution for the risk of this process exists for the general initial state.
Therefore, we calculate the risk numerically.
These calculations are computationally intensive due to the $2^{m \times M}$ possible measurement outcomes upon which the risk depends.

Figure \ref{fig:mm} plots the Holevo and $1$-$0$ risks as a function of the resources used for $1 \leq M \leq 5$ when the QPEA register is prepared in the optimal cosine state.
For noiseless circuits (top row), the Holevo risk scales as $\mathcal{R}_{\mathcal{Q}}$ for any fixed $M$.
The minimum risk per resource occurs when $M=1$.
Therefore, the QPEA circuit should be run once in this situation.
The same is true for the absolute and squared risks in noiseless systems (not plotted here).
For the $1$-$0$ loss in noiseless circuits, better risk scaling is seen for larger $M$.
Risk below $\mathcal{R}_{\mathcal{C}}$ is achieved for $M \geq 3$.
\new{One concludes that the QPEA circuit with a register prepared in the optimal cosine state should be measured at least three times.}
When depolarizing noise is present (bottom row of Fig. \ref{fig:mm}), both risks tend to $L_0$ as register size increases for any $M$.
The minimum risk obtained over all register sizes decreases as $M$ increases. 
For large enough $M$, risks below $\mathcal{R}_{\mathcal{C}}$ can be achieved by fixing register size and $M$ appropriately.

\section{Conclusions}

We have derived  the optimal input state for the QPEA with respect to Bayesian risk functions. The state, we showed, is  equivalent to the eigenvector of a Toeplitz matrix with the smallest eigenvalue.
The elements of this matrix are the Fourier coefficients of the target loss function that is used to calculate the risk of interest.
Although a closed form of these states may not exist in general, we find that a good approximation exists in the form of a cosine state.
Optimizing the frequency of the  cosine state enables theoretically optimal noiseless risk scalings for three common risks.
In addition, we demonstrate that the traditional QPEA state never minimizes risk for any non-constant loss function.
We have also shown that if any depolarizing noise exists, the risk will tend to a constant in the large-resource limit. This is regardless of the preparation of the register state.
Nevertheless, the noisy scalings can be improved by repeating measurements of a circuit with a fixed register size $m$ a number of times $M$.
Data for small $m \times M$ shows that performance can be improved beyond classical shot-noise risk limits.
However, these improvements are minor when compared to modern algorithms based on adaptive and iterative repetitive measurements of non-entangled circuits \cite{me, me2}.

\bibliography{paper}

\begin{thebibliography}{33}%
\makeatletter
\providecommand \@ifxundefined [1]{%
 \@ifx{#1\undefined}
}%
\providecommand \@ifnum [1]{%
 \ifnum #1\expandafter \@firstoftwo
 \else \expandafter \@secondoftwo
 \fi
}%
\providecommand \@ifx [1]{%
 \ifx #1\expandafter \@firstoftwo
 \else \expandafter \@secondoftwo
 \fi
}%
\providecommand \natexlab [1]{#1}%
\providecommand \enquote  [1]{``#1''}%
\providecommand \bibnamefont  [1]{#1}%
\providecommand \bibfnamefont [1]{#1}%
\providecommand \citenamefont [1]{#1}%
\providecommand \href@noop [0]{\@secondoftwo}%
\providecommand \href [0]{\begingroup \@sanitize@url \@href}%
\providecommand \@href[1]{\@@startlink{#1}\@@href}%
\providecommand \@@href[1]{\endgroup#1\@@endlink}%
\providecommand \@sanitize@url [0]{\catcode `\\12\catcode `\$12\catcode `\&12\catcode `\#12\catcode `\^12\catcode `\_12\catcode `\%12\relax}%
\providecommand \@@startlink[1]{}%
\providecommand \@@endlink[0]{}%
\providecommand \url  [0]{\begingroup\@sanitize@url \@url }%
\providecommand \@url [1]{\endgroup\@href {#1}{\urlprefix }}%
\providecommand \urlprefix  [0]{URL }%
\providecommand \Eprint [0]{\href }%
\providecommand \doibase [0]{https://doi.org/}%
\providecommand \selectlanguage [0]{\@gobble}%
\providecommand \bibinfo  [0]{\@secondoftwo}%
\providecommand \bibfield  [0]{\@secondoftwo}%
\providecommand \translation [1]{[#1]}%
\providecommand \BibitemOpen [0]{}%
\providecommand \bibitemStop [0]{}%
\providecommand \bibitemNoStop [0]{.\EOS\space}%
\providecommand \EOS [0]{\spacefactor3000\relax}%
\providecommand \BibitemShut  [1]{\csname bibitem#1\endcsname}%
\let\auto@bib@innerbib\@empty
%</preamble>
\bibitem [{\citenamefont {Brassard}\ \emph {et~al.}(2002)\citenamefont {Brassard}, \citenamefont {Hoyer}, \citenamefont {Mosca},\ and\ \citenamefont {Tapp}}]{QAE}%
  \BibitemOpen
  \bibfield  {author} {\bibinfo {author} {\bibfnamefont {G.}~\bibnamefont {Brassard}}, \bibinfo {author} {\bibfnamefont {P.}~\bibnamefont {Hoyer}}, \bibinfo {author} {\bibfnamefont {M.}~\bibnamefont {Mosca}},\ and\ \bibinfo {author} {\bibfnamefont {A.}~\bibnamefont {Tapp}},\ }\href@noop {} {\bibfield  {journal} {\bibinfo  {journal} {Contemporary Mathematics}\ }\textbf {\bibinfo {volume} {305}},\ \bibinfo {pages} {53} (\bibinfo {year} {2002})}\BibitemShut {NoStop}%
\bibitem [{\citenamefont {Suzuki}\ \emph {et~al.}(2020)\citenamefont {Suzuki}, \citenamefont {Uno}, \citenamefont {Raymond}, \citenamefont {Tanaka}, \citenamefont {Onodera},\ and\ \citenamefont {Yamamoto}}]{QAE2}%
  \BibitemOpen
  \bibfield  {author} {\bibinfo {author} {\bibfnamefont {Y.}~\bibnamefont {Suzuki}}, \bibinfo {author} {\bibfnamefont {S.}~\bibnamefont {Uno}}, \bibinfo {author} {\bibfnamefont {R.}~\bibnamefont {Raymond}}, \bibinfo {author} {\bibfnamefont {T.}~\bibnamefont {Tanaka}}, \bibinfo {author} {\bibfnamefont {T.}~\bibnamefont {Onodera}},\ and\ \bibinfo {author} {\bibfnamefont {N.}~\bibnamefont {Yamamoto}},\ }\href@noop {} {\bibfield  {journal} {\bibinfo  {journal} {Quantum Inf. Process.}\ }\textbf {\bibinfo {volume} {19}},\ \bibinfo {pages} {1} (\bibinfo {year} {2020})}\BibitemShut {NoStop}%
\bibitem [{\citenamefont {Lloyd}\ \emph {et~al.}(2020)\citenamefont {Lloyd}, \citenamefont {Bosch}, \citenamefont {Palma}, \citenamefont {Kiani}, \citenamefont {Liu}, \citenamefont {Marvian}, \citenamefont {Rebentrost},\ and\ \citenamefont {Arvidsson-Shukur}}]{dave2}%
  \BibitemOpen
  \bibfield  {author} {\bibinfo {author} {\bibfnamefont {S.}~\bibnamefont {Lloyd}}, \bibinfo {author} {\bibfnamefont {S.}~\bibnamefont {Bosch}}, \bibinfo {author} {\bibfnamefont {G.~D.}\ \bibnamefont {Palma}}, \bibinfo {author} {\bibfnamefont {B.}~\bibnamefont {Kiani}}, \bibinfo {author} {\bibfnamefont {Z.-W.}\ \bibnamefont {Liu}}, \bibinfo {author} {\bibfnamefont {M.}~\bibnamefont {Marvian}}, \bibinfo {author} {\bibfnamefont {P.}~\bibnamefont {Rebentrost}},\ and\ \bibinfo {author} {\bibfnamefont {D.~M.}\ \bibnamefont {Arvidsson-Shukur}},\ }\href@noop {} {} (\bibinfo {year} {2020}),\ \Eprint {https://arxiv.org/abs/2006.00841} {arXiv:2006.00841} \BibitemShut {NoStop}%
\bibitem [{\citenamefont {Shor}(1994)}]{shor}%
  \BibitemOpen
  \bibfield  {author} {\bibinfo {author} {\bibfnamefont {P.}~\bibnamefont {Shor}},\ }\bibfield  {title} {\bibinfo {title} {Algorithms for quantum computation: discrete logarithms and factoring},\ }in\ \href {https://doi.org/10.1109/SFCS.1994.365700} {\emph {\bibinfo {booktitle} {Proc. of the 35th FOCS}}}\ (\bibinfo {year} {IEEE, New York, 1994})\ pp.\ \bibinfo {pages} {124--134}\BibitemShut {NoStop}%
\bibitem [{\citenamefont {Lloyd}\ \emph {et~al.}(2021)\citenamefont {Lloyd}, \citenamefont {Kiani}, \citenamefont {Arvidsson-Shukur}, \citenamefont {Bosch}, \citenamefont {Palma}, \citenamefont {Kaminsky}, \citenamefont {Liu},\ and\ \citenamefont {Marvian}}]{dave1}%
  \BibitemOpen
  \bibfield  {author} {\bibinfo {author} {\bibfnamefont {S.}~\bibnamefont {Lloyd}}, \bibinfo {author} {\bibfnamefont {B.~T.}\ \bibnamefont {Kiani}}, \bibinfo {author} {\bibfnamefont {D.~R.~M.}\ \bibnamefont {Arvidsson-Shukur}}, \bibinfo {author} {\bibfnamefont {S.}~\bibnamefont {Bosch}}, \bibinfo {author} {\bibfnamefont {G.~D.}\ \bibnamefont {Palma}}, \bibinfo {author} {\bibfnamefont {W.~M.}\ \bibnamefont {Kaminsky}}, \bibinfo {author} {\bibfnamefont {Z.-W.}\ \bibnamefont {Liu}},\ and\ \bibinfo {author} {\bibfnamefont {M.}~\bibnamefont {Marvian}},\ }\href@noop {} {} (\bibinfo {year} {2021}),\ \Eprint {https://arxiv.org/abs/2104.01410} {arXiv:2104.01410} \BibitemShut {NoStop}%
\bibitem [{\citenamefont {Bauer}\ \emph {et~al.}(2020)\citenamefont {Bauer}, \citenamefont {Bravyi}, \citenamefont {Motta},\ and\ \citenamefont {Chan}}]{algos1}%
  \BibitemOpen
  \bibfield  {author} {\bibinfo {author} {\bibfnamefont {B.}~\bibnamefont {Bauer}}, \bibinfo {author} {\bibfnamefont {S.}~\bibnamefont {Bravyi}}, \bibinfo {author} {\bibfnamefont {M.}~\bibnamefont {Motta}},\ and\ \bibinfo {author} {\bibfnamefont {G.~K.-L.}\ \bibnamefont {Chan}},\ }\href@noop {} {\bibfield  {journal} {\bibinfo  {journal} {Chem. Rev.}\ }\textbf {\bibinfo {volume} {120}},\ \bibinfo {pages} {12685} (\bibinfo {year} {2020})}\BibitemShut {NoStop}%
\bibitem [{\citenamefont {Whitfield}\ \emph {et~al.}(2011)\citenamefont {Whitfield}, \citenamefont {Biamonte},\ and\ \citenamefont {Aspuru-Guzik}}]{VQE}%
  \BibitemOpen
  \bibfield  {author} {\bibinfo {author} {\bibfnamefont {J.~D.}\ \bibnamefont {Whitfield}}, \bibinfo {author} {\bibfnamefont {J.}~\bibnamefont {Biamonte}},\ and\ \bibinfo {author} {\bibfnamefont {A.}~\bibnamefont {Aspuru-Guzik}},\ }\href@noop {} {\bibfield  {journal} {\bibinfo  {journal} {Molecular Physics}\ }\textbf {\bibinfo {volume} {109}},\ \bibinfo {pages} {735} (\bibinfo {year} {2011})}\BibitemShut {NoStop}%
\bibitem [{\citenamefont {Lin}\ and\ \citenamefont {Tong}(2022)}]{VQE2}%
  \BibitemOpen
  \bibfield  {author} {\bibinfo {author} {\bibfnamefont {L.}~\bibnamefont {Lin}}\ and\ \bibinfo {author} {\bibfnamefont {Y.}~\bibnamefont {Tong}},\ }\href@noop {} {\bibfield  {journal} {\bibinfo  {journal} {Phys. Rev. X Quantum}\ }\textbf {\bibinfo {volume} {3}},\ \bibinfo {pages} {010318} (\bibinfo {year} {2022})}\BibitemShut {NoStop}%
\bibitem [{\citenamefont {Demkowicz-Dobrzanski}\ \emph {et~al.}(2009)\citenamefont {Demkowicz-Dobrzanski}, \citenamefont {Dorner}, \citenamefont {Smith}, \citenamefont {Lundeen}, \citenamefont {Wasilewski}, \citenamefont {Banaszek},\ and\ \citenamefont {Walmsley}}]{inter2}%
  \BibitemOpen
  \bibfield  {author} {\bibinfo {author} {\bibfnamefont {R.}~\bibnamefont {Demkowicz-Dobrzanski}}, \bibinfo {author} {\bibfnamefont {U.}~\bibnamefont {Dorner}}, \bibinfo {author} {\bibfnamefont {B.~J.}\ \bibnamefont {Smith}}, \bibinfo {author} {\bibfnamefont {J.~S.}\ \bibnamefont {Lundeen}}, \bibinfo {author} {\bibfnamefont {W.}~\bibnamefont {Wasilewski}}, \bibinfo {author} {\bibfnamefont {K.}~\bibnamefont {Banaszek}},\ and\ \bibinfo {author} {\bibfnamefont {I.~A.}\ \bibnamefont {Walmsley}},\ }\href {https://doi.org/10.1103/PhysRevA.80.013825} {\bibfield  {journal} {\bibinfo  {journal} {Phys. Rev. A}\ }\textbf {\bibinfo {volume} {80}},\ \bibinfo {pages} {013825} (\bibinfo {year} {2009})}\BibitemShut {NoStop}%
\bibitem [{\citenamefont {Oblak}\ \emph {et~al.}(2005)\citenamefont {Oblak}, \citenamefont {Petrov}, \citenamefont {Alzar}, \citenamefont {Tittel}, \citenamefont {Vershovski}, \citenamefont {Mikkelsen}, \citenamefont {S{\o}rensen},\ and\ \citenamefont {Polzik}}]{clocks}%
  \BibitemOpen
  \bibfield  {author} {\bibinfo {author} {\bibfnamefont {D.}~\bibnamefont {Oblak}}, \bibinfo {author} {\bibfnamefont {P.~G.}\ \bibnamefont {Petrov}}, \bibinfo {author} {\bibfnamefont {C.~L.~G.}\ \bibnamefont {Alzar}}, \bibinfo {author} {\bibfnamefont {W.}~\bibnamefont {Tittel}}, \bibinfo {author} {\bibfnamefont {A.~K.}\ \bibnamefont {Vershovski}}, \bibinfo {author} {\bibfnamefont {J.~K.}\ \bibnamefont {Mikkelsen}}, \bibinfo {author} {\bibfnamefont {J.~L.}\ \bibnamefont {S{\o}rensen}},\ and\ \bibinfo {author} {\bibfnamefont {E.~S.}\ \bibnamefont {Polzik}},\ }\href@noop {} {\bibfield  {journal} {\bibinfo  {journal} {Phys. Rev. A}\ }\textbf {\bibinfo {volume} {71}},\ \bibinfo {pages} {043807} (\bibinfo {year} {2005})}\BibitemShut {NoStop}%
\bibitem [{\citenamefont {Barish}\ and\ \citenamefont {Weiss}(1999)}]{grav_waves}%
  \BibitemOpen
  \bibfield  {author} {\bibinfo {author} {\bibfnamefont {B.~C.}\ \bibnamefont {Barish}}\ and\ \bibinfo {author} {\bibfnamefont {R.}~\bibnamefont {Weiss}},\ }\href@noop {} {\bibfield  {journal} {\bibinfo  {journal} {Phys. Today}\ }\textbf {\bibinfo {volume} {52}},\ \bibinfo {pages} {44} (\bibinfo {year} {1999})}\BibitemShut {NoStop}%
\bibitem [{\citenamefont {Giovannetti}\ \emph {et~al.}(2011)\citenamefont {Giovannetti}, \citenamefont {Lloyd},\ and\ \citenamefont {Maccone}}]{Lloyd1}%
  \BibitemOpen
  \bibfield  {author} {\bibinfo {author} {\bibfnamefont {V.}~\bibnamefont {Giovannetti}}, \bibinfo {author} {\bibfnamefont {S.}~\bibnamefont {Lloyd}},\ and\ \bibinfo {author} {\bibfnamefont {L.}~\bibnamefont {Maccone}},\ }\href {https://doi.org/10.1038/nphoton.2011.35} {\bibfield  {journal} {\bibinfo  {journal} {Nat. Photonics}\ }\textbf {\bibinfo {volume} {5}},\ \bibinfo {pages} {222–229} (\bibinfo {year} {2011})}\BibitemShut {NoStop}%
\bibitem [{\citenamefont {{Giovannetti}}\ \emph {et~al.}(2006)\citenamefont {{Giovannetti}}, \citenamefont {{Lloyd}},\ and\ \citenamefont {{Maccone}}}]{Lloyd2}%
  \BibitemOpen
  \bibfield  {author} {\bibinfo {author} {\bibfnamefont {V.}~\bibnamefont {{Giovannetti}}}, \bibinfo {author} {\bibfnamefont {S.}~\bibnamefont {{Lloyd}}},\ and\ \bibinfo {author} {\bibfnamefont {L.}~\bibnamefont {{Maccone}}},\ }\href {https://doi.org/10.1103/PhysRevLett.96.010401} {\bibfield  {journal} {\bibinfo  {journal} {\prl}\ }\textbf {\bibinfo {volume} {96}},\ \bibinfo {eid} {010401} (\bibinfo {year} {2006})}\BibitemShut {NoStop}%
\bibitem [{\citenamefont {Smith}\ \emph {et~al.}(2022)\citenamefont {Smith}, \citenamefont {Barnes},\ and\ \citenamefont {Arvidsson-Shukur}}]{me}%
  \BibitemOpen
  \bibfield  {author} {\bibinfo {author} {\bibfnamefont {J.~G.}\ \bibnamefont {Smith}}, \bibinfo {author} {\bibfnamefont {C.~H.~W.}\ \bibnamefont {Barnes}},\ and\ \bibinfo {author} {\bibfnamefont {D.~R.~M.}\ \bibnamefont {Arvidsson-Shukur}},\ }\href {https://doi.org/10.1103/PhysRevA.106.062615} {\bibfield  {journal} {\bibinfo  {journal} {Phys. Rev. A}\ }\textbf {\bibinfo {volume} {106}},\ \bibinfo {pages} {062615} (\bibinfo {year} {2022})}\BibitemShut {NoStop}%
\bibitem [{\citenamefont {Demkowicz-Dobrza{\'n}ski}\ \emph {et~al.}(2012)\citenamefont {Demkowicz-Dobrza{\'n}ski}, \citenamefont {Ko{\l}ody{\'n}ski},\ and\ \citenamefont {Gu{\c{t}}{\u{a}}}}]{noisy_QFI}%
  \BibitemOpen
  \bibfield  {author} {\bibinfo {author} {\bibfnamefont {R.}~\bibnamefont {Demkowicz-Dobrza{\'n}ski}}, \bibinfo {author} {\bibfnamefont {J.}~\bibnamefont {Ko{\l}ody{\'n}ski}},\ and\ \bibinfo {author} {\bibfnamefont {M.}~\bibnamefont {Gu{\c{t}}{\u{a}}}},\ }\href@noop {} {\bibfield  {journal} {\bibinfo  {journal} {Nat. Commun.}\ }\textbf {\bibinfo {volume} {3}},\ \bibinfo {pages} {1063} (\bibinfo {year} {2012})}\BibitemShut {NoStop}%
\bibitem [{\citenamefont {Albarelli}\ and\ \citenamefont {Demkowicz-Dobrza\ifmmode~\acute{n}\else \'{n}\fi{}ski}(2022)}]{noisy_QFI2}%
  \BibitemOpen
  \bibfield  {author} {\bibinfo {author} {\bibfnamefont {F.}~\bibnamefont {Albarelli}}\ and\ \bibinfo {author} {\bibfnamefont {R.}~\bibnamefont {Demkowicz-Dobrza\ifmmode~\acute{n}\else \'{n}\fi{}ski}},\ }\href {https://doi.org/10.1103/PhysRevX.12.011039} {\bibfield  {journal} {\bibinfo  {journal} {Phys. Rev. X}\ }\textbf {\bibinfo {volume} {12}},\ \bibinfo {pages} {011039} (\bibinfo {year} {2022})}\BibitemShut {NoStop}%
\bibitem [{\citenamefont {Smith}\ \emph {et~al.}(2024)\citenamefont {Smith}, \citenamefont {Barnes},\ and\ \citenamefont {Arvidsson-Shukur}}]{me2}%
  \BibitemOpen
  \bibfield  {author} {\bibinfo {author} {\bibfnamefont {J.~G.}\ \bibnamefont {Smith}}, \bibinfo {author} {\bibfnamefont {C.~H.}\ \bibnamefont {Barnes}},\ and\ \bibinfo {author} {\bibfnamefont {D.~R.}\ \bibnamefont {Arvidsson-Shukur}},\ }\href@noop {} {\bibfield  {journal} {\bibinfo  {journal} {Phys. Rev. A}\ }\textbf {\bibinfo {volume} {109}},\ \bibinfo {pages} {042412} (\bibinfo {year} {2024})}\BibitemShut {NoStop}%
\bibitem [{\citenamefont {Berry}\ \emph {et~al.}(2009)\citenamefont {Berry}, \citenamefont {Higgins}, \citenamefont {Bartlett}, \citenamefont {Mitchell}, \citenamefont {Pryde},\ and\ \citenamefont {Wiseman}}]{berry}%
  \BibitemOpen
  \bibfield  {author} {\bibinfo {author} {\bibfnamefont {D.~W.}\ \bibnamefont {Berry}}, \bibinfo {author} {\bibfnamefont {B.~L.}\ \bibnamefont {Higgins}}, \bibinfo {author} {\bibfnamefont {S.~D.}\ \bibnamefont {Bartlett}}, \bibinfo {author} {\bibfnamefont {M.~W.}\ \bibnamefont {Mitchell}}, \bibinfo {author} {\bibfnamefont {G.~J.}\ \bibnamefont {Pryde}},\ and\ \bibinfo {author} {\bibfnamefont {H.~M.}\ \bibnamefont {Wiseman}},\ }\href@noop {} {\bibfield  {journal} {\bibinfo  {journal} {Phys. Rev. A}\ }\textbf {\bibinfo {volume} {80}},\ \bibinfo {pages} {052114} (\bibinfo {year} {2009})}\BibitemShut {NoStop}%
\bibitem [{\citenamefont {Svore}\ \emph {et~al.}(2013)\citenamefont {Svore}, \citenamefont {Hastings},\ and\ \citenamefont {Freedman}}]{fourier}%
  \BibitemOpen
  \bibfield  {author} {\bibinfo {author} {\bibfnamefont {K.~M.}\ \bibnamefont {Svore}}, \bibinfo {author} {\bibfnamefont {M.}~\bibnamefont {Hastings}},\ and\ \bibinfo {author} {\bibfnamefont {M.}~\bibnamefont {Freedman}},\ }\href@noop {} {\bibfield  {journal} {\bibinfo  {journal} {Quantum Inf. Comput.}\ }\textbf {\bibinfo {volume} {14}},\ \bibinfo {pages} {306} (\bibinfo {year} {2013})}\BibitemShut {NoStop}%
\bibitem [{\citenamefont {O’Brien}\ \emph {et~al.}(2019)\citenamefont {O’Brien}, \citenamefont {Tarasinski},\ and\ \citenamefont {Terhal}}]{fourier2}%
  \BibitemOpen
  \bibfield  {author} {\bibinfo {author} {\bibfnamefont {T.~E.}\ \bibnamefont {O’Brien}}, \bibinfo {author} {\bibfnamefont {B.}~\bibnamefont {Tarasinski}},\ and\ \bibinfo {author} {\bibfnamefont {B.~M.}\ \bibnamefont {Terhal}},\ }\href@noop {} {\bibfield  {journal} {\bibinfo  {journal} {New J. Phys.}\ }\textbf {\bibinfo {volume} {21}},\ \bibinfo {pages} {023022} (\bibinfo {year} {2019})}\BibitemShut {NoStop}%
\bibitem [{\citenamefont {Humphreys}\ \emph {et~al.}(2013)\citenamefont {Humphreys}, \citenamefont {Barbieri}, \citenamefont {Datta},\ and\ \citenamefont {Walmsley}}]{multiphase}%
  \BibitemOpen
  \bibfield  {author} {\bibinfo {author} {\bibfnamefont {P.~C.}\ \bibnamefont {Humphreys}}, \bibinfo {author} {\bibfnamefont {M.}~\bibnamefont {Barbieri}}, \bibinfo {author} {\bibfnamefont {A.}~\bibnamefont {Datta}},\ and\ \bibinfo {author} {\bibfnamefont {I.~A.}\ \bibnamefont {Walmsley}},\ }\href@noop {} {\bibfield  {journal} {\bibinfo  {journal} {Phys. Rev. Lett.}\ }\textbf {\bibinfo {volume} {111}},\ \bibinfo {pages} {070403} (\bibinfo {year} {2013})}\BibitemShut {NoStop}%
\bibitem [{\citenamefont {Nielsen}\ and\ \citenamefont {Chuang}(2000)}]{nielsen}%
  \BibitemOpen
  \bibfield  {author} {\bibinfo {author} {\bibfnamefont {M.~A.}\ \bibnamefont {Nielsen}}\ and\ \bibinfo {author} {\bibfnamefont {I.}~\bibnamefont {Chuang}},\ }\href@noop {} {\emph {\bibinfo {title} {Quantum computation and quantum information}}}\ (\bibinfo  {publisher} {Cambridge University Press},\ \bibinfo {year} {2000})\BibitemShut {NoStop}%
\bibitem [{\citenamefont {Griffiths}\ and\ \citenamefont {Niu}(1996)}]{semiQFT}%
  \BibitemOpen
  \bibfield  {author} {\bibinfo {author} {\bibfnamefont {R.~B.}\ \bibnamefont {Griffiths}}\ and\ \bibinfo {author} {\bibfnamefont {C.-S.}\ \bibnamefont {Niu}},\ }\href@noop {} {\bibfield  {journal} {\bibinfo  {journal} {Phys. Rev. Let.}\ }\textbf {\bibinfo {volume} {76}},\ \bibinfo {pages} {3228} (\bibinfo {year} {1996})}\BibitemShut {NoStop}%
\bibitem [{\citenamefont {Chiaverini}\ \emph {et~al.}(2005)\citenamefont {Chiaverini}, \citenamefont {Britton}, \citenamefont {Leibfried}, \citenamefont {Knill}, \citenamefont {Barrett}, \citenamefont {Blakestad}, \citenamefont {Itano}, \citenamefont {Jost}, \citenamefont {Langer}, \citenamefont {Ozeri}, \citenamefont {Schaetz},\ and\ \citenamefont {Wineland}}]{semiQFT2}%
  \BibitemOpen
  \bibfield  {author} {\bibinfo {author} {\bibfnamefont {J.~a.}\ \bibnamefont {Chiaverini}}, \bibinfo {author} {\bibfnamefont {J.}~\bibnamefont {Britton}}, \bibinfo {author} {\bibfnamefont {D.}~\bibnamefont {Leibfried}}, \bibinfo {author} {\bibfnamefont {E.}~\bibnamefont {Knill}}, \bibinfo {author} {\bibfnamefont {M.~D.}\ \bibnamefont {Barrett}}, \bibinfo {author} {\bibfnamefont {R.}~\bibnamefont {Blakestad}}, \bibinfo {author} {\bibfnamefont {W.~M.}\ \bibnamefont {Itano}}, \bibinfo {author} {\bibfnamefont {J.~D.}\ \bibnamefont {Jost}}, \bibinfo {author} {\bibfnamefont {C.}~\bibnamefont {Langer}}, \bibinfo {author} {\bibfnamefont {R.}~\bibnamefont {Ozeri}}, \bibinfo {author} {\bibfnamefont {T.}~\bibnamefont {Schaetz}},\ and\ \bibinfo {author} {\bibfnamefont {D.~J.}\ \bibnamefont {Wineland}},\ }\href@noop {} {\bibfield  {journal} {\bibinfo  {journal} {science}\ }\textbf {\bibinfo {volume} {308}},\ \bibinfo {pages} {997} (\bibinfo {year} {2005})}\BibitemShut {NoStop}%
\bibitem [{\citenamefont {{Kitaev}}(1995)}]{QPEA}%
  \BibitemOpen
  \bibfield  {author} {\bibinfo {author} {\bibfnamefont {A.~Y.}\ \bibnamefont {{Kitaev}}},\ }\href@noop {} {} (\bibinfo {year} {1995}),\ \Eprint {https://arxiv.org/abs/arXiv:quant-ph/9511026} {arXiv:quant-ph/9511026} \BibitemShut {NoStop}%
\bibitem [{\citenamefont {Lu}\ and\ \citenamefont {Lin}(2022)}]{QPEA2}%
  \BibitemOpen
  \bibfield  {author} {\bibinfo {author} {\bibfnamefont {X.}~\bibnamefont {Lu}}\ and\ \bibinfo {author} {\bibfnamefont {H.}~\bibnamefont {Lin}},\ }\href@noop {} {} (\bibinfo {year} {2022}),\ \Eprint {https://arxiv.org/abs/arXiv:2210.00231} {arXiv:2210.00231} \BibitemShut {NoStop}%
\bibitem [{\citenamefont {Chapeau-Blondeau}\ and\ \citenamefont {Belin}(2020)}]{QPEA_vs_MLE}%
  \BibitemOpen
  \bibfield  {author} {\bibinfo {author} {\bibfnamefont {F.}~\bibnamefont {Chapeau-Blondeau}}\ and\ \bibinfo {author} {\bibfnamefont {E.}~\bibnamefont {Belin}},\ }\href {https://doi.org/10.1007/s12243-020-00803-1} {\bibfield  {journal} {\bibinfo  {journal} {Ann. Telecommun.}\ }\textbf {\bibinfo {volume} {75}},\ \bibinfo {pages} {641–653} (\bibinfo {year} {2020})}\BibitemShut {NoStop}%
\bibitem [{\citenamefont {Higgins}\ \emph {et~al.}(2007)\citenamefont {Higgins}, \citenamefont {Berry}, \citenamefont {Bartlett}, \citenamefont {Wiseman},\ and\ \citenamefont {Pryde}}]{higgins}%
  \BibitemOpen
  \bibfield  {author} {\bibinfo {author} {\bibfnamefont {B.~L.}\ \bibnamefont {Higgins}}, \bibinfo {author} {\bibfnamefont {D.~W.}\ \bibnamefont {Berry}}, \bibinfo {author} {\bibfnamefont {S.~D.}\ \bibnamefont {Bartlett}}, \bibinfo {author} {\bibfnamefont {H.~M.}\ \bibnamefont {Wiseman}},\ and\ \bibinfo {author} {\bibfnamefont {G.~J.}\ \bibnamefont {Pryde}},\ }\href@noop {} {\bibfield  {journal} {\bibinfo  {journal} {Nature}\ }\textbf {\bibinfo {volume} {450}},\ \bibinfo {pages} {393} (\bibinfo {year} {2007})}\BibitemShut {NoStop}%
\bibitem [{\citenamefont {van Dam}\ \emph {et~al.}(2007)\citenamefont {van Dam}, \citenamefont {D’Ariano}, \citenamefont {Ekert}, \citenamefont {Macchiavello},\ and\ \citenamefont {Mosca}}]{opt_states}%
  \BibitemOpen
  \bibfield  {author} {\bibinfo {author} {\bibfnamefont {W.}~\bibnamefont {van Dam}}, \bibinfo {author} {\bibfnamefont {G.~M.}\ \bibnamefont {D’Ariano}}, \bibinfo {author} {\bibfnamefont {A.}~\bibnamefont {Ekert}}, \bibinfo {author} {\bibfnamefont {C.}~\bibnamefont {Macchiavello}},\ and\ \bibinfo {author} {\bibfnamefont {M.}~\bibnamefont {Mosca}},\ }\href@noop {} {\bibfield  {journal} {\bibinfo  {journal} {Phys. Rev. Lett.}\ }\textbf {\bibinfo {volume} {98}},\ \bibinfo {pages} {090501} (\bibinfo {year} {2007})}\BibitemShut {NoStop}%
\bibitem [{\citenamefont {Rendon}\ \emph {et~al.}(2022)\citenamefont {Rendon}, \citenamefont {Izubuchi},\ and\ \citenamefont {Kikuchi}}]{cosine_windows}%
  \BibitemOpen
  \bibfield  {author} {\bibinfo {author} {\bibfnamefont {G.}~\bibnamefont {Rendon}}, \bibinfo {author} {\bibfnamefont {T.}~\bibnamefont {Izubuchi}},\ and\ \bibinfo {author} {\bibfnamefont {Y.}~\bibnamefont {Kikuchi}},\ }\href@noop {} {\bibfield  {journal} {\bibinfo  {journal} {Phys. Rev. D}\ }\textbf {\bibinfo {volume} {106}},\ \bibinfo {pages} {034503} (\bibinfo {year} {2022})}\BibitemShut {NoStop}%
\bibitem [{\citenamefont {Ekstr{\"o}m}\ \emph {et~al.}(2018)\citenamefont {Ekstr{\"o}m}, \citenamefont {Garoni},\ and\ \citenamefont {Serra-Capizzano}}]{toeplitz2}%
  \BibitemOpen
  \bibfield  {author} {\bibinfo {author} {\bibfnamefont {S.-E.}\ \bibnamefont {Ekstr{\"o}m}}, \bibinfo {author} {\bibfnamefont {C.}~\bibnamefont {Garoni}},\ and\ \bibinfo {author} {\bibfnamefont {S.}~\bibnamefont {Serra-Capizzano}},\ }\href@noop {} {\bibfield  {journal} {\bibinfo  {journal} {Exp. Math.}\ }\textbf {\bibinfo {volume} {27}},\ \bibinfo {pages} {478} (\bibinfo {year} {2018})}\BibitemShut {NoStop}%
\bibitem [{\citenamefont {Ekstr{\"o}m}\ and\ \citenamefont {Serra-Capizzano}(2018)}]{toeplitz}%
  \BibitemOpen
  \bibfield  {author} {\bibinfo {author} {\bibfnamefont {S.-E.}\ \bibnamefont {Ekstr{\"o}m}}\ and\ \bibinfo {author} {\bibfnamefont {S.}~\bibnamefont {Serra-Capizzano}},\ }\href@noop {} {\bibfield  {journal} {\bibinfo  {journal} {Numer. Linear Algebra Appl.}\ }\textbf {\bibinfo {volume} {25}},\ \bibinfo {pages} {e2137} (\bibinfo {year} {2018})}\BibitemShut {NoStop}%
\bibitem [{\citenamefont {Harrow}\ \emph {et~al.}(2009)\citenamefont {Harrow}, \citenamefont {Hassidim},\ and\ \citenamefont {Lloyd}}]{hhl}%
  \BibitemOpen
  \bibfield  {author} {\bibinfo {author} {\bibfnamefont {A.~W.}\ \bibnamefont {Harrow}}, \bibinfo {author} {\bibfnamefont {A.}~\bibnamefont {Hassidim}},\ and\ \bibinfo {author} {\bibfnamefont {S.}~\bibnamefont {Lloyd}},\ }\href@noop {} {\bibfield  {journal} {\bibinfo  {journal} {Phys. Rev. Lett.}\ }\textbf {\bibinfo {volume} {103}},\ \bibinfo {pages} {150502} (\bibinfo {year} {2009})}\BibitemShut {NoStop}%
\end{thebibliography}%

\end{document}